\documentclass[12pt,preprint]{aastex}
\usepackage{graphicx}
\usepackage{float}
\usepackage{amsmath}
\usepackage{epsfig,floatflt}

\newcommand{\Npix}{N_{\rm pix}}
\newcommand{\Opix}{\Omega_{\rm pix}}

\newcommand{\nside}{N_{\rm side}}

\newcommand{\hpx}{HEALPix }
\newcommand{\healpixns}{\textbf{HEALPix}}
\renewcommand{\mod}{{\ \rm mod \ }}

\newcommand{\beq}{\begin{equation}}
\newcommand{\eeq}{\end{equation}}

\newcommand{\ntheta}{N_{\theta}}
\newcommand{\nphi}{N_{\phi}}
\newcommand{\integ}{{\rm I}}
\newcommand{\frow}{f_{\rm row}}
\newcommand{\thetapix}{\theta_{\rm pix}}
 
\begin{document}

 
\title{HEALPix --- a Framework for High Resolution Discretization, and Fast Analysis of Data Distributed on the Sphere}

\author{K.\ M.\ G\'orski\altaffilmark{{1,2}},
        E.\ Hivon\altaffilmark{{3,4}},
        A.\ J.\ Banday\altaffilmark{5},
        B.\ D.\ Wandelt\altaffilmark{{6,7}},
        F.\ K.\ Hansen\altaffilmark{8},
        M.\ Reinecke\altaffilmark{5},
        M. Bartelman\altaffilmark{9}
       }

\affil{
$^{1}$ JPL/Caltech, M/S 169/327, 4800 Oak Grove Drive, Pasadena CA 91109 \\
$^{2}$ Warsaw University Observatory, Aleje Ujazdowskie 4, 00-478 Warszawa, Poland \\
$^{3}$ Observational Cosmology, MS 59-33, Caltech, Pasadena, CA 91125 \\
$^{4}$ IPAC, MS 100-22, Caltech, Pasadena, CA 91125 \\
$^{5}$ Max-Planck-Institut f\"ur Astrophysik, Karl-Schwarzschild-Str.\\
       1, Postfach 1317,\\D-85741 Garching bei M\"unchen, Germany \\
$^{6}$ Department of Physics, University of Illinois, Urbana, IL 61801 \\
$^{7}$ Department of Astronomy, UIUC, 1002 W. Green Street, Urbana, IL 61801 \\
$^{8}$ Dipartimento di Fisica, Universita di Roma 'Tor Vergata', Via della
Ricerca Scientifica 1, I-00133 Roma, Italy \\
$^{9}$ ITA, Universit\"at Heidelberg, Tiergartenstr. 15, D-69121 Heidelberg, Germany\\       
      }


\begin{abstract}

HEALPix -- the Hierarchical Equal Area iso-Latitude Pixelization --
is a versatile data structure with an associated library of
computational algorithms and visualization software that supports fast
scientific applications executable directly on very large volumes of
astronomical data and large area surveys in the form of discretized spherical
maps. Originally developed to address the data processing and analysis needs of
the present generation of cosmic microwave background (CMB) experiments
(e.g. BOOMERanG, WMAP), HEALPix
can be expanded to meet many of the profound challenges that will arise in
confrontation with the observational output of future missions and
experiments, including e.g. Planck, Herschel, SAFIR, and the Beyond Einstein CMB
polarization probe. 
In this paper we consider the requirements and constraints to be met
in order to implement a sufficient framework for 
the efficient discretization and fast analysis/synthesis of functions defined on the sphere,
and summarise how they are satisfied by HEALPix. 

\end{abstract}

\section{Introduction}

Advanced detectors in modern astronomy produce data at huge rates at many
wavelengths. Some data sets are becoming very impressively large indeed. Of
particular interest to us are those that accumulate astronomical data
distributed on the entire sky, or a considerable fraction thereof. Typical
examples include radio, cosmic microwave background, submillimeter, infrared,
X-ray, and gamma-ray sky maps of diffuse emission, and full sky or wide area
surveys of extragalactic objects.  Together with the wealth of gathered
information comes the inevitable burden of increased complexity of the tasks of
data reduction and science extraction. In this paper we are focused on those
issues, which are related to the distinctive nature of the spherical spatial
domain over which the data reside, and need to be analysed for scientific
return. Our original motivations come from work in the field of measurement and
interpretation of the cosmic microwave background (CMB) anisotropy. The growing
complexity of the CMB anisotropy science extraction problem can be illustrated
by the transition between the data sets of COBE-DMR (early 1990s, 7 deg
resolution, ~6000 pixel sky maps at 3 wavelengths), Boomerang (late 1990s,
FWHM~12 arcmin, partial sky maps of ~200000 pixels at 4 wavelengths), WMAP
(early 2000s, resolution up to FWHM~14 arcmin, ~3 million pixel sky maps at 5
wavelengths), and Planck (data expected ~2008, resolution up to FWHM~5 arcmin,
~50 million pixel sky maps at 9 wavelengths).  Science extraction from these
data sets involves (1) global analysis problems -- harmonic decomposition,
estimation of the power spectrum and higher order measures of spatial
correlations, (2) simulations of models of the primary and foreground sky
signals to study instrument performance, and calibrate foreground separation and
statistical inference methods, (3) real space morphological analyses, object
detection, identification, and characterization, and (4) spatial and/or spectral
cross-correlation with external data sets. These tasks, and many others,
necessitate a careful definition of the data models, and proper set-up of the
mathematical framework of data analysis such that the algorithmic and computing
time requirements can be satisfied in order to achieve successful scientific
interpretation of expensive and precious observations. A particular method of
addressing some of these problems is described next.

\section{Discretized Mapping and Analysis of Functions on the Sphere}

The analysis of functions on  domains with spherical topology occupies a
central place in physical science and engineering disciplines.
This is particularly apparent in the fields of astronomy, cosmology,
geophysics,  atomic and nuclear physics. In many cases the geometry is either
dictated by the object under study or approximate spherical symmetry can be
exploited to yield powerful perturbation methods. Practical
limits for the purely analytical study of these problems create
an urgent necessity for efficient and accurate numerical tools.
 
The simplicity of the spherical form belies the intricacy of global
analysis on the sphere. There is no known
point set which achieves the analogue of uniform sampling in Euclidean space and
allows exact and invertible discrete spherical harmonic decompositions
of arbitrary but band-limited functions. Any existing proposition of practical
schemes for the  discrete treatment of such functions
on the sphere  introduces some (hopefully small)
systematic error dependent on the global properties of
the point set. The goal is to minimise these errors and
faithfully represent deterministic functions as well as realizations of
random variates both
in configuration and Fourier space while maintaining computational efficiency.

We illustrate these points using as an example the field of Cosmic Microwave Background (CMB)
anisotropies. Here we are already in a situation of an ongoing  rapid growth of the volume 
of the available data.
The  NASA's Wilkinson Microwave Anisotropy Probe (WMAP) already is,
and ESA's Planck will (in the near future) be aiming
to provide multi-frequency, high resolution, full sky measurements of the anisotropy in
both temperature and polarization of the cosmic microwave background radiation.
The ultimate data products of these missions ---
multiple microwave sky maps, each of which will have to comprise
more than $\sim 10^6$ pixels in order to render the angular
resolution of the instruments ---
will present serious challenges to those involved in the
analysis and scientific exploitation of the results of both surveys.

A digitized sky map is an essential intermediate
stage in information processing between
the entry point of data acquisition by the
instruments --- very large time ordered data streams,
and the final stage of astrophysical analysis ---
typically producing a $\lq$few' numerical values
of physical parameters of interest.
{\it COBE}-DMR sky maps (angular resolution of $7^\circ$ (FWHM) in
three frequency bands, two channels each, 6144 pixels per map)
were considered large at the time of their release.
As for the presently available (WMAP) and future CMB maps, a whole sky CMB survey
at the angular resolution
of $\sim 10'$ (FWHM), discretized with
a few pixels per resolution element
(so that the discretisation effects on the signal are
sub-dominant with respect to the effects of instrument's angular response),
will require map sizes of at least
$N_{pix}\sim $ a few $\times 1.5\, 10^6$ pixels.
Even more pixels than that will be needed to represent the Planck-HFI higher
resolution high-frequency channels.

This estimate, $N_{pix}$, should be multiplied by  the number of frequency bands
(or, indeed, by the number of individual
observing channels --- $74$ in the case of Planck --- for the analysis work
to be done before the
final coadded maps are made for each frequency band) to render
an approximate expected
size of the already very compressed form of survey data which would
be the input to the astrophysical analysis pipeline.

Hence, a careful
attention ought to be given to devising  high resolution CMB map
structures which can maximally facilitate
the forthcoming analyses of large size data sets, especially so because
many essential scientific questions
can only be answered by {\em global} studies of  such data sets.

This paper describes the essential geometric and algebraic properties 
of our method of digital 
representation of  functions on the sphere
 --- the Hierarchical
Equal Area and isoLatitude Pixelization (HEALPix) ---
and the associated multi-purpose computer
software package. We have originally devised \hpx, and we started 
distributing \hpx software to the community in 1997. Presently \hpx
software is distributed from the web-site www.eso.org/science/healpix.

\section{Requirements for a Spherical Pixelization Scheme}

Numerical analysis of functions on the sphere involves
(1) a class of mathematical operations, whose objects are
(2) discretised maps, i.e. quantizations of arbitrary functions
according to a chosen tessellation (exhaustive partition of the sphere into
finite area elements). Hereafter we mostly specialise our discussion
to CMB related applications of
\hpx, but all our statements hold true generally for any relevant
deterministic and random functions on the sphere.

Considering point (1):
Standard operations of numerical analysis which one might wish to
execute on the sphere include
convolutions with local and global kernels,
Fourier analysis with spherical harmonics
and power spectrum estimation,
wavelet decomposition, nearest-neighbour searches, topological
analysis, including searches for extrema or zero-crossings,
computing Minkowski functionals,
extraction of patches and
finite differencing for solving partial
differential equations.
Some of these operations become prohibitively slow
if the sampling of functions on the sphere, and the related structure of
the discrete data set, are not designed carefully.

Regarding point (2):
Typically, a whole sky map rendered by a CMB experiment contains
({\it i}) signals coming from the sky,
which are by design strongly band-width limited (in the sense of
spatial Fourier
decomposition) by the instrument's angular response
function, and
({\it ii}) a projection into the elements of a discrete map, or pixels,
of the observing instrument's noise; this pixel noise should be random,
and white, at least near the discretisation scale, with a band-width
significantly exceeding that of all the signals.

With these considerations in mind we proposed the following list of desiderata 
for the mathematical structure of discrete whole sky maps:

\begin{enumerate} 

\item {\bf Hierarchical structure of the database}. This is recognized
as essential for very large data bases, and was indeed postulated
already in construction of the Quadrilateralized Spherical Cube
(or QuadCube, see \citet{QuadCube:1992}, and \hfill\break  http://lambda.gsfc.nasa.gov/product/cobe/skymap\_info\_new.cfm),
which was used for all the {\it COBE} data.
A simple argument in favour of this states that the data elements
which are nearby in a multi-dimensional configuration space
(here, on the surface of a sphere), are also nearby in the tree
structure of the data base. This property facilitates various
topological methods of analysis, and allows for easy construction
of wavelet transforms on triangular and quadrilateral grids
through fast look-up of nearest
neighbors. 

\item {\bf Equal areas of discrete elements of partition}. This is advantageous because
white noise at the sampling frequency of the instrument gets integrated exactly into
white noise in the pixel space, and sky signals are sampled without regional dependence
(although still care must be taken to choose a pixel size
sufficiently small compared to the
instrumental resolution to avoid excessive, and pixel shape dependent, signal
smoothing).

\item {\bf Iso-Latitude distribution of discrete area elements on a sphere}.  This property
is essential for computational speed in all operations involving evaluations of spherical
harmonics. Since the associated Legendre polynomials are evaluated via
slow recursions, any sampling grid deviations from an iso-latitude
distribution result in a prohibitive loss of computational performance
with the growing number of sampling points.

\end{enumerate}

Various known sampling distributions on a sphere 
have been used for the discretisation and analysis
of functions (for example, see Driscoll \& Healy (1994),
Muciaccia, Natoli \& Vittorio (1998) --- rectangular grids,
Baumgardner \& Frederickson (1985),  Tegmark (1996) --- icosahedral grids,
Saff \& Kuijlaars (1997),  Crittenden \& Turok (1998) ---  'igloo' grids,
and Szalay \& Brunner (1998)  --- a triangular grid), but each
fails to meet simultaneously all of these requirements. In particular,

i) Quadrilateralized Spherical Cube obeys points 1 and
(appro\-xi\-ma\-te\-ly) 2, but fails on point 3,
and cannot be used for efficient Fourier analysis at high resolution.
 
ii) Equidistant Cylindrical Projection, a very common computational
tool in geophysics,
and climate modeling, and recently implemented for CMB work
(Muciaccia, Natoli, Vittorio, 1998), satisfies points 1 and 3,
but by construction
fails with point 2. This is a nuisance from the point of view of
application to full sky survey data due to wasteful oversampling
near the poles of the map (or - while angular resolution of the measurements 
is fixed by the instrument and does not vary over the sky, the
map resolution, or pixel size, depends on the distance form the poles). 

iii) Hexagonal sampling grids with icosahedral symmetry
perform  superbly in those applications where near uniformity
of sampling on a sphere is essential (Saff, Kuijlaars, 1997),
and can be devised to meet requirement 2 (see e.g. Tegmark 1996).
However, by construction they fail {\it both} points 1 and 3.

iv) Igloo-type constructions are devised to satisfy point 3
(E. Wright, 1997, private communication; Crittenden \& Turok, 1998).
Point 2 can be satisfied to reasonable accuracy if quite a large
number of base-resolution pixels is used( which, however,  precludes
efficient construction of simple wavelet transforms).
Conversely, a tree-structure seeded with a small number of
base-resolution pixels forces significant variations in both
area and shape of the pixels.

v) GLESP construction \citep{GLESP:2003} takes advantage of the Gauss-Legendre
quadrature to render high accuracy of numerical integration, but allows irregular 
variations the in pixel area and is not hierarchical --- in fact it offers no relation between
the tessellations derived at different resolutions.

\section{Meeting the Requirements}

All the requirements introduced in section 3. are satisfied by the class of spherical tessellations
structured as follows. 

\begin{figure}[p]
\centerline{\psfig{figure=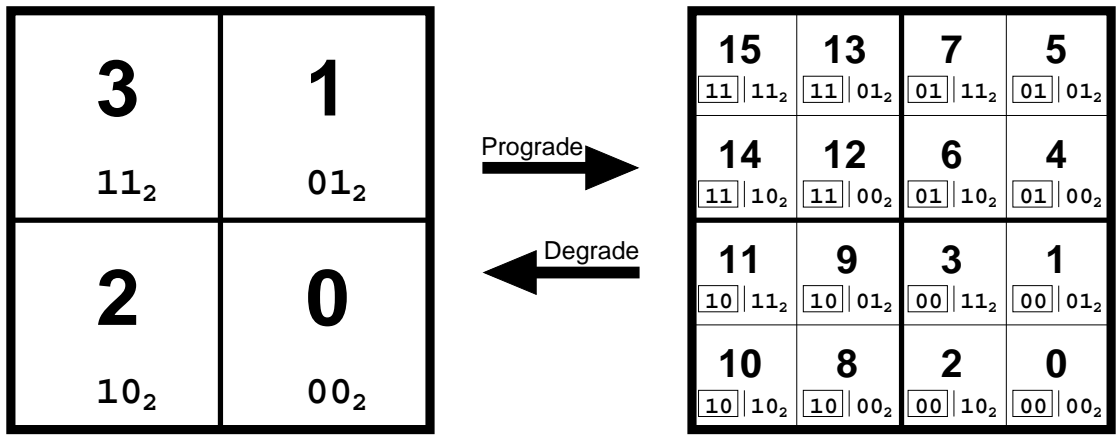,width=0.9\textwidth}}
\caption
{Quadrilateral tree pixel numbering.
The coarsely pixelised coordinate patch on
the left consists
of four pixels. Two bits suffice to label the pixels.
To increase the resolution, every
pixel splits into
4 daughter pixels shown on the right. These daughters inherit the pixel
index of their
parent (boxed) and acquire
two new bits to give the new pixel index.
Several such curvilinearly mapped coordinate patches
(12 in the case of \hpx, and 6 in the case of the {\it COBE} quad-sphere)
are joined at the boundaries to cover
the sphere. All pixels indices carry a prefix (here omitted for clarity)
which identifies which base-resolution pixel they belong to.
\label{fig:quadtree}}
\end{figure}

First  let us assume that the sphere is partitioned into a number of curvilinear quadrilaterals,
which constitute the base level tessellation.
If there exists a mapping of each element of partition onto a square $[0,1]\times[0,1]$, 
a nested $n\times n$ subdivision of the square into ever diminishing sub-elements obtains trivially, and
a hierarchical tree structure of resulting database follows. For example, a $2\times 2$ partition renders
a quadrilateral tree, which admits an elegant binary indexation (illustrated in Fig. \ref{fig:quadtree})
previously employed in construction
of the QuadCube spherical pixelization. 

\begin{figure*}[p]
\centering\mbox{\includegraphics[width=1.0\textwidth,clip]{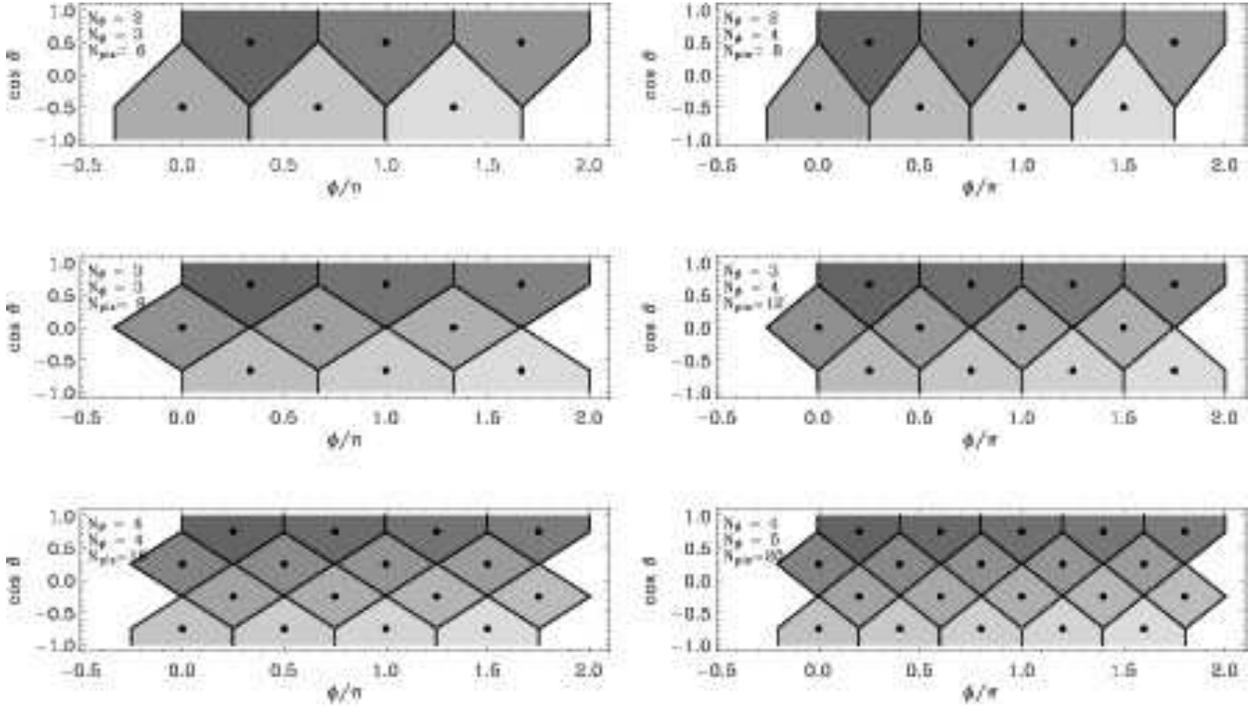}}
\caption[]{cylindrical projection of a number of equal area, iso-latitude
tessellations of the sphere, which can support a hierarchical tree of further subdivisions
of each large base-resolution pixel. Six out of possible realizations of such tessellation are shown for
several values of grid parameters $N_\theta$, and $N_\phi$.
The HEALPix grid, used for development of the corresponding software package, 
is described by $N_\theta=3$, and $N_\phi=4$.
}
\label{fig:multi_cyll}
\end{figure*}

\begin{figure*}[p]
\centering\mbox{\includegraphics[width=0.6\textwidth,clip]{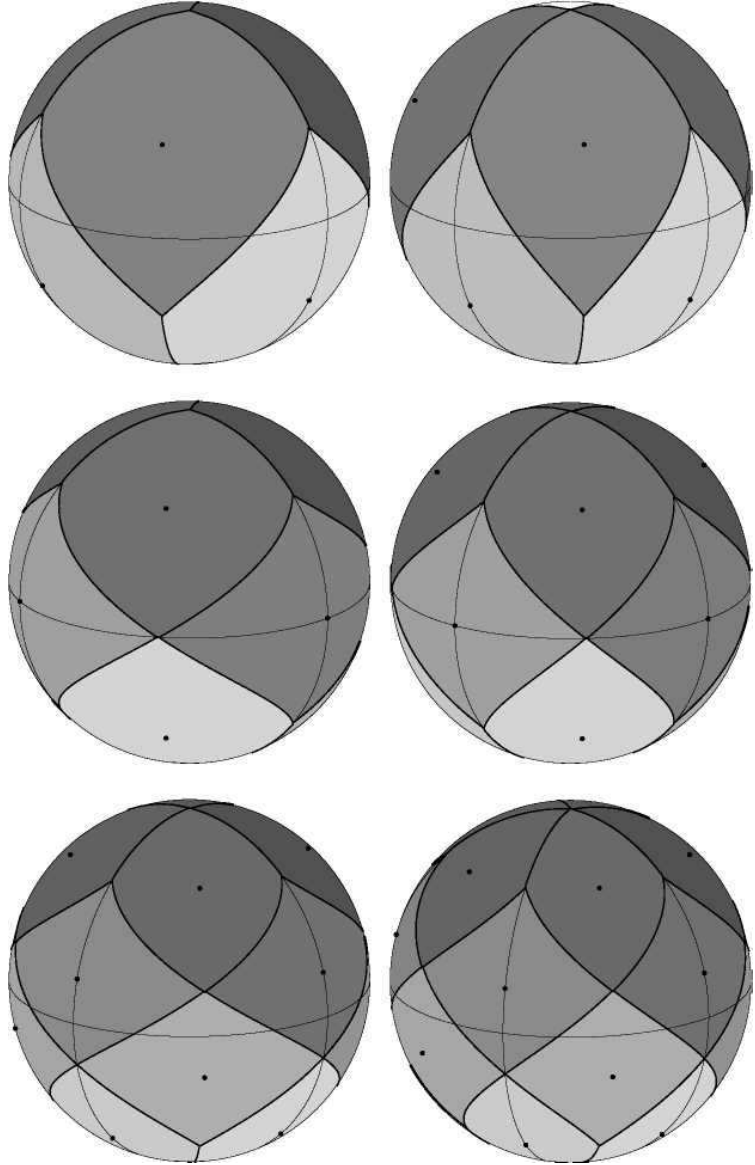}}
\caption[]{Orthographic projection of the same base-resolution spherical tessellations as those shown in Fig.\ref{fig:multi_cyll}.
}
\label{fig:multi_orth}
\end{figure*}

Next, let us consider the base level spherical tessellation. An entire class of such tessellations
can be constructed as illustrated in Figs. \ref{fig:multi_cyll}, and \ref{fig:multi_orth}.
These constructions are characterized by two parameters:
$N_\theta$ - a number of the base-resolution pixel layers between the north and south poles, and
$N_\phi$ - a multiplicity of the meridional cuts, or the number of equatorial, or circum-polar
base-resolutions pixels. Obviously, the total number of base-resolution pixels is equal to
$N_{pix} = N_\theta \times N_\phi $, and the area of each one of them is equal to
$\Omega_{pix} = 4\pi/N_\theta/ N_\phi$.
One may also notice (see Fig. \ref{fig:multi_cyll}) that each tessellation
includes two single layers of polar cup pixels (with or without an azimuthal twist in their respective
positions on the sphere for odd, and even values of $N_\theta$, respectively), and an
$N_\theta - 2$ layers of equatorial zone pixels, which form a regular rhomboidal grid in the
cylindrical projection of the sphere.    Since the cylindrical projection is an area preserving mapping,
this property immediately illustrates that the areas of equatorial zone pixels are all equal, and to meet
our requirement of fully equal area partition of the sphere, we need to demonstrate that our constructions
render identical areas of the polar pixels as well. Indeed, this allows to formulate a constraint
on the latitude $\theta_\star$ at which the lateral vertices of both polar and equatorial pixels meet:
\begin{equation}
2\pi\times (1-\cos \theta_\star) = N_\phi \times \Omega_{pix}/2, \quad {\rm hence} \quad \cos \theta_\star = 
(N_\theta -1)/N_\theta.
\end{equation}

The curvilinear quadrilateral pixels of this tessellation class retain equal areas, but vary in shape
depending on their positions on the sphere. We have chosen the $N_\theta = 3$, and $N_\phi=4$
grid (middle row, right column in both Figs. \ref{fig:multi_cyll}, and \ref{fig:multi_orth}) 
for definition of our digital full sky map data standard, and named it the
HEALPix grid (G\'orski et al. 1998). This grid is the basis for development of the HEALPix software, 
and it is
described in more detail in the next section.

\section{The HEALPix Grid}

\begin{figure*}[p]
\centerline{\psfig{figure=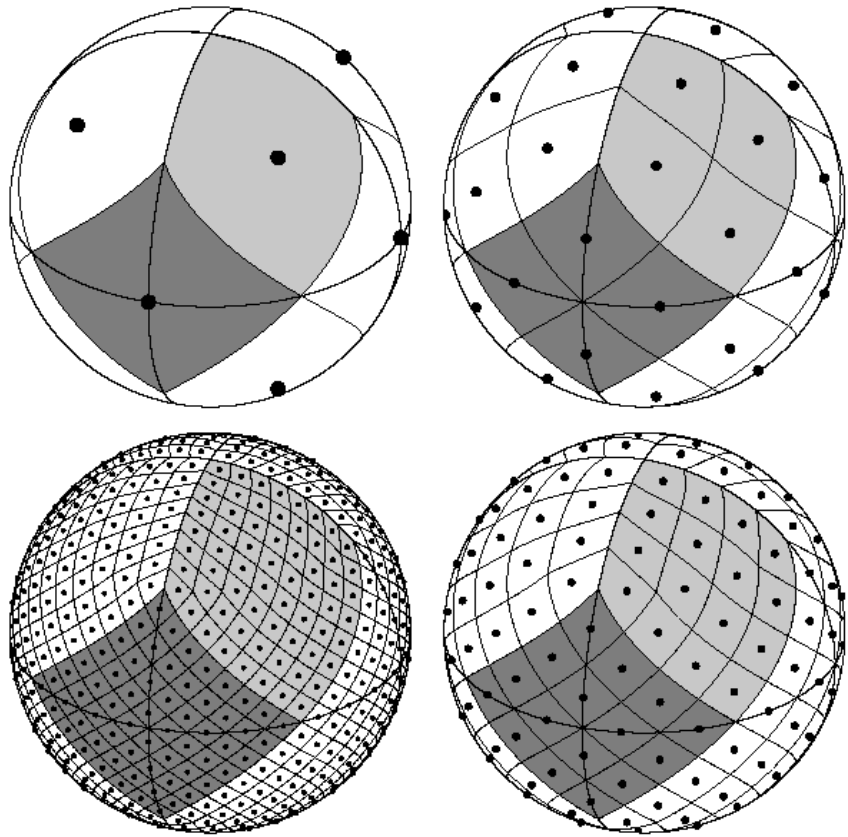,width=0.7\textwidth,bbllx=0pt,bblly=100pt,bburx=610pt,bbury=700pt}}
\caption[]{Orthographic view of the \hpx ~partition of the sphere.
Overplot of equator and  meridians illustrates the octahedral symmetry of
\hpx.
Light-gray shading shows one of the eight (four north, and four south)
identical polar
base-resolution pixels.
Dark-gray shading shows one of the four identical equatorial
base-resolution pixels.
Moving clockwise from the upper left
panel the grid is hierarchically subdivided with
the grid resolution parameter equal to $N_{side} = \,1,\,2,\,4,\,8$,
and the total number of pixels  equal to
$N_{pix} = 12 \times N_{side}^2 = \,12,\,48,\,192,\,768$.
All pixel centers are located on $N_{ring} = 4 \times N_{side} - 1$ rings of
constant latitude.
Within each panel the areas of all pixels are identical.}
\label{fig:HEALPix}
\end{figure*}

The requirements formulated in Sec.3 are satisfied by construction with the
Hierarchical Equal Area, iso-Latitude Pixelisation (\healpixns)
of the sphere, which is shown in Figure (\ref{fig:HEALPix}).
This Figure demonstrates that
HEALPix is built geometrically as a self-similar,
refinable quadrilateral mesh on the sphere.
The base-resolution comprises twelve pixels in three
rings around the poles and equator.
The resolution of the grid is expressed by parameter $N_{side}$ which
defines the number of divisions along the side of a base-resolution
pixel that is needed to reach a desired high-resolution partition.
All pixel centers are placed on rings of constant latitude,
and are equidistant in azimuth
(on each ring). All iso-latitude rings located between the upper and
lower corners of the equatorial base-resolution pixels
(i.e. $-2/3 < \cos \theta_\star < 2/3$), or in the
equatorial zone, are divided into the same number of pixels:
$N_{eq}= 4\times N_{side}$. The remaining rings are located within the
polar cap regions
($\vert \cos \theta_\star \vert > 2/3$)
 and contain a varying number of pixels, increasing
from ring to ring, with increasing distance
from the poles, by one pixel within each quadrant.

A \hpx map has $\Npix=12\nside^2$ pixels of the same area $\Opix = \frac{
\pi}{3\nside^2}$.

\subsection{Pixel Positions}

Locations on the sphere are  defined by $(z\equiv\cos\theta,\phi)$ where 
$\theta \in [0,\pi]$ is the colatitude in radians measured from the 
North Pole and $\phi \in [0,2\pi]$ is the longitude in radians measured Eastward.

For a resolution parameter $\nside$, the pixels are layed out on $4\nside-1$
iso-latitude rings, and can be ordered using the pixel index
$p \in [0,\Npix]$ running around those rings from the North to South pole.

Pixel centers on the northern hemisphere are given by the following equations:

North polar cap --- for $p_h=(p+1)/2$, the  ring index $1\leq i < \nside$, and the pixel-in-ring index $1 \leq j \leq 4i$
\begin{eqnarray}
i &=& I(\sqrt{p_h-\sqrt{I(p_h)}}) + 1\\
j &=& p+1-2i(i-1)\\
z &=& 1 - \frac{i^2}{3\nside^2} \label{eq:pixncapz}\\
\phi &=& \frac{\pi}{2i} \left(j-\frac{s}{2}\right), \quad {\rm and} \quad s=1 \label{eq:pixncapphi}
\end{eqnarray}

North equatorial belt --- for  $p'=p-2\nside(\nside-1)$, $\nside \leq i \leq 2\nside$, and $1 \leq j \leq 4\nside$
\begin{eqnarray}
i&=&I(p'/4\nside)+\nside\\
j&=&(p' \mod 4\nside) +1\\
z &=& \frac{4}{3} - \frac{2i}{3\nside} \label{eq:pixeqbeltz}\\
\phi &=& \frac{\pi}{2\nside}\left(j -\frac{s}{2}\right),
	\quad {\rm and}\quad s = (i-\nside+1)\mod 2, \label{eq:pixeqbeltphi}
\end{eqnarray}
where the auxilliary index $s$ describes the phase shifts along the rings, and
$\integ(x)$ is the largest integer number smaller than $x$.

Pixel center positions on the Southern hemisphere are obtained by 
the mirror symmetry of the grid with respect to the Equator ($z=0$).

Defining $\Delta z$ and $\Delta \phi$ as the variation of $z$ and $\phi$ when
$i$ and $j$ are respectively increased by unity, one can check that discretized
area element
$\left|\Delta z \Delta \phi \right| = \Opix$, i.e. it is a constant.

\subsection{Pixel Indexing}

\begin{figure}[p]
\includegraphics[%
  bb=160pt 70pt 377pt 737pt, clip,angle=90, width=0.95\textwidth]%
  {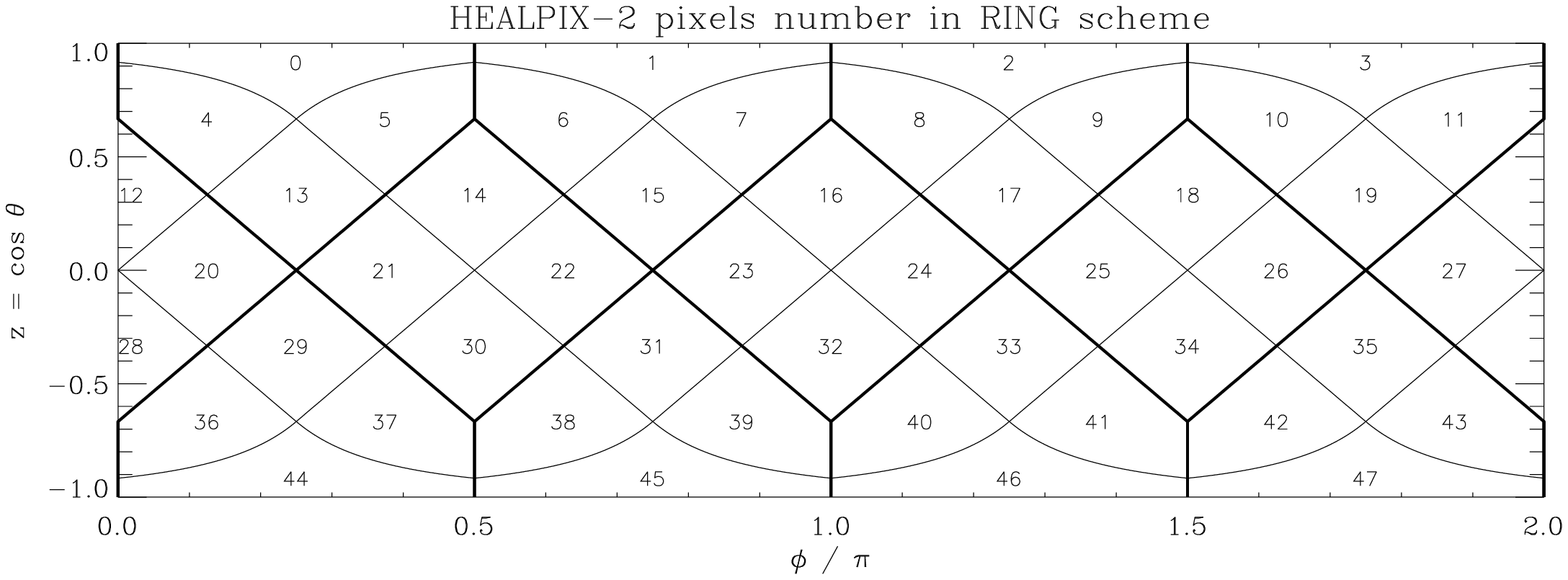}
\includegraphics[%
  bb=160pt 70pt 377pt 737pt, clip,angle=90, width=0.95\textwidth]%
  {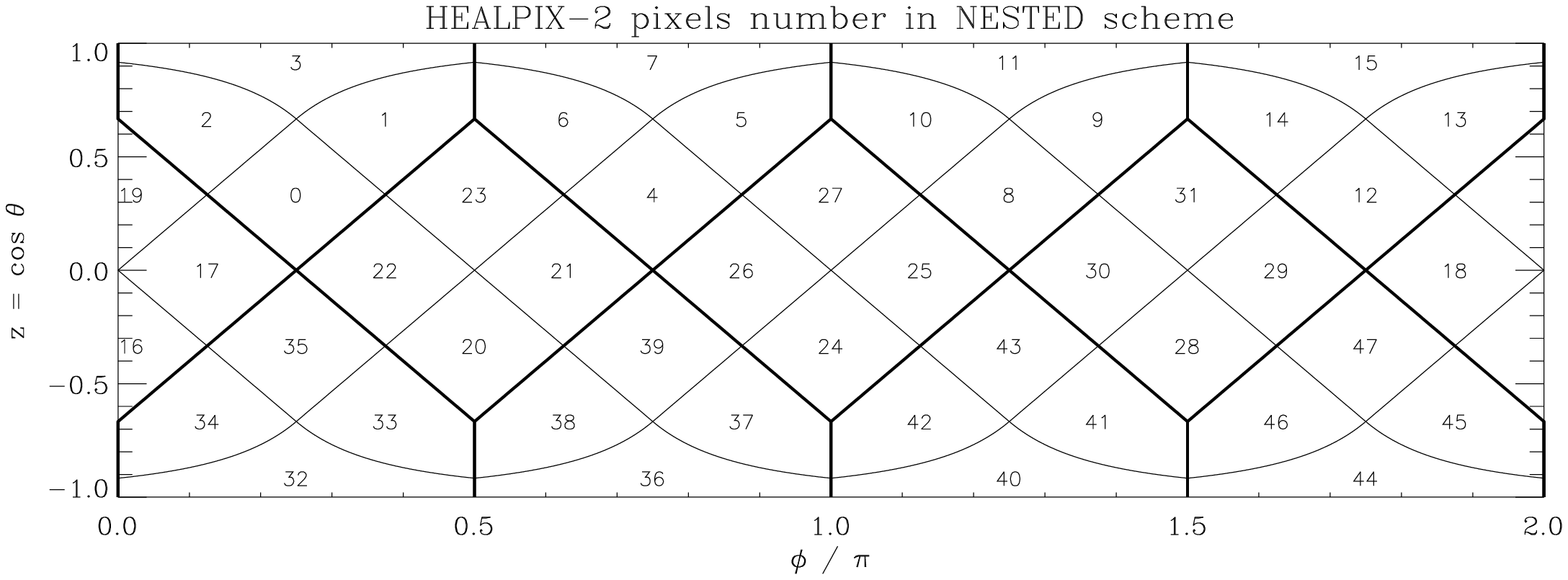}
\includegraphics[%
  bb=160pt 70pt 377pt 737pt, clip,angle=90, width=0.95\textwidth]%
  {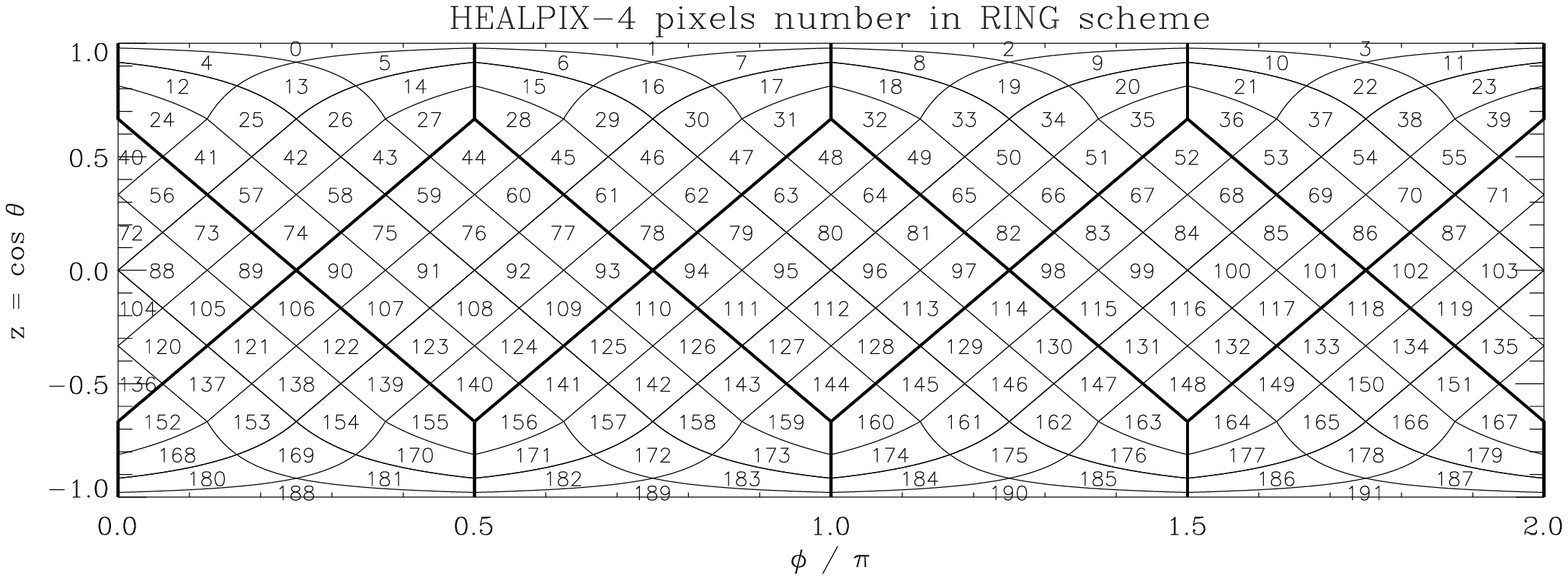}
\includegraphics[%
  bb=160pt 70pt 377pt 737pt, clip,angle=90, width=0.95\textwidth]%
  {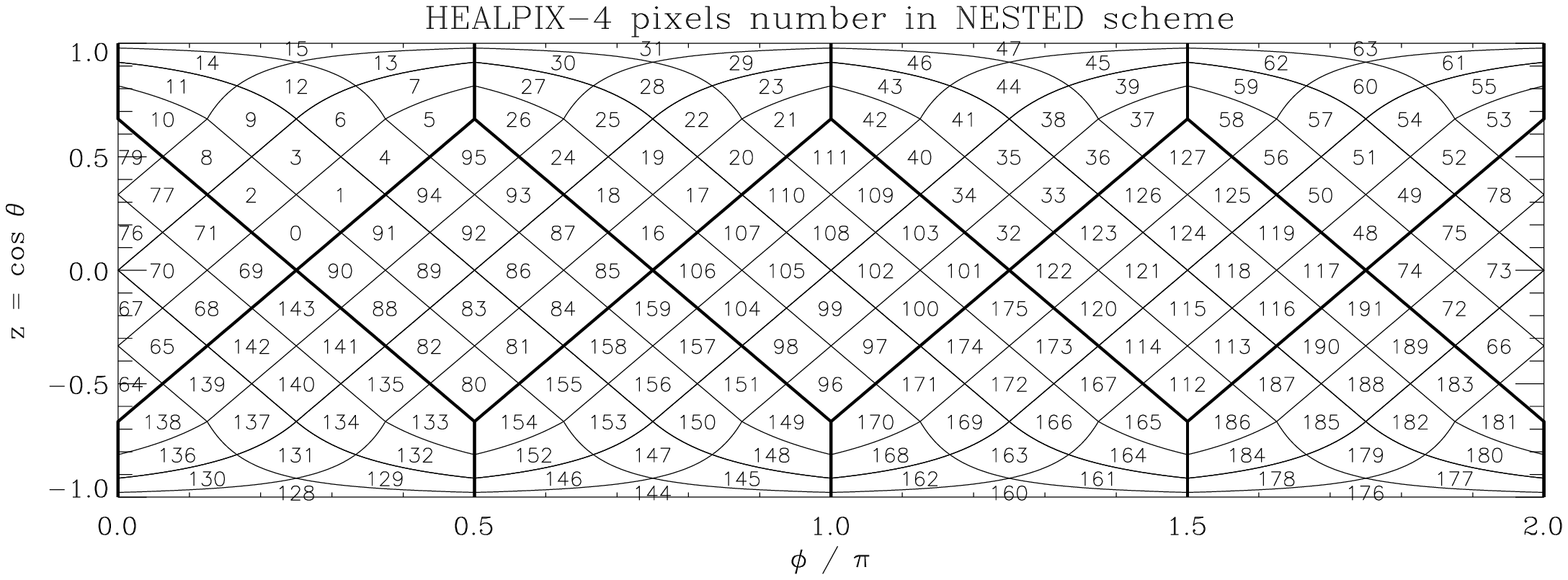}
\caption{The layout of the HEALPix pixels on the sphere, and demonstration of two possible 
pixel indexetions --- one running on iso-latitude rings, and the other arranged hierarchically, or
in a nested tree fashion. The latter capability is essential for the possible
database applications of HEALPix.}
\label{fig:cylproj}
\end{figure}

Specific geometrical properties allow HEALPix to support two different
numbering schemes for the pixels, as illustrated in the Figure (\ref{fig:cylproj}).

First, one can simply count the pixels moving down from the north
to the south pole along each iso-latitude ring.
It is in this scheme that Fourier transforms with spherical harmonics
are easy to implement.

Second, one can replicate the tree structure of pixel numbering used
e.g. with the Quadrilateralized Spherical Cube. This can easily be
implemented since, due to the simple description of pixel boundaries,
the analytical mapping of the \hpx ~base-resolution elements
(curvilinear quadrilaterals) into a [0,1]$\times$[0,1] square exists.
This tree structure, a.k.a. nested scheme, allows one to implement
efficiently all applications involving  nearest-neighbour searches
(see Wandelt, Hivon, and G\'orski 1998),
and also allows for an immediate
construction of the fast Haar wavelet transform on \hpx.

The base-resolution pixel index number $f$ runs in $\left\{0,\ntheta\nphi-1\right\}=\{0,11\}$. 
Introducing the row index 
\begin{equation}
	\frow = \integ(f/\nphi) \label{eq:rowindex}
\end{equation}
we define
two functions which index the location of the southernmost corner (or vertex) of
each base resolution pixel on the sphere, in latitude and longitude respectively:
\begin{eqnarray}
	F_1(f) &=& \frow + 2, \label{eq:F1face} \\
	F_2(f) &=& 2 (f \mod \nphi) - (\frow \mod 2) + 1. \label{eq:F2face}
\end{eqnarray}

Consider the nested index $p_n\in [0, 12\nside^2-1]$, and define $p'_n = (p_n \mod \nside^2)$, where $p'_n$
denotes the nested pixel index within each base-resolution element., which has the following 
binary representation $p'_n=[\overline{\dots b_2 b_1 b_0}]_2$ (and $b_i = 0$,  or $1$, and has the weight $2^i$).

Given a grid resolution parameter $\nside$ the location of a pixel center on each base-resolution pixel 
is represented by the 2 indices $x$ and $y$  $\in \{0,\nside-1\}$. They both have
their origin in the southernmost corner of each base-resolution pixel, with the $x$ index running
along the North-East direction, while the $y$ index runs along the North-West direction.
The binary representation of $p'_n$ determines the values of $x$, and $y$ as the following combinations of 
even and odd bits, respectively
\begin{eqnarray}
x&=&[\overline{\dots b_2  b_0}]_2\\
y&=&[\overline{\dots b_3  b_1}]_2
\end{eqnarray}
Next, we introduce the vertical and horizontal coordinates (within the base-resolution pixel)
\begin{eqnarray}
	v&=&x+y \\
	h&=&x-y 
\end{eqnarray}
and obtain the following relations for the ring index $i$ ($\in \{1, (\ntheta+1)\nside-1\}$) 
\begin{equation}
	i = F_1(f) \nside - v - 1,
\end{equation}
and the longitude index 
\begin{equation}
	j = \frac{F_2(f) \nside + h + s }{2},
\end{equation}
which can be translated into ($z,\phi$) coordinates using Eqs. \ref{eq:pixncapz},
\ref{eq:pixncapphi},
\ref{eq:pixeqbeltz}, and
\ref{eq:pixeqbeltphi}.

\subsection{Pixel Boundaries}

Pixel boundaries are non-geodesic and take a very simple
form: $\cos \theta = a + b \times \phi$ in the equatorial zone,
and $\cos \theta = a + b / \phi^2$ in the polar caps.
This allows one to explicitly check by simple analytical integration the
exact area equality among pixels,
and to compute efficiently more complex objects,
e.g. the Fourier transforms of individual pixels.

Since the pixel center location is parametrized by integer value of $j$, setting $j=k+1/2$
or $j=k+1/2+i$ with $k$ a positive integer in Eq.~(\ref{eq:pixncapphi}) and substituting into
Eq.~(\ref{eq:pixncapz}) will give for the pixels boundaries of the North polar
cap ($z > 2/3$)
\begin{eqnarray}
z &=& 1 - \frac{k^2}{3\nside^2}\left(\frac{\pi}{2\phi_t}\right)^2, \\
z &=& 1 - \frac{k^2}{3\nside^2}\left(\frac{\pi}{2\phi_t-\pi}\right)^2,
\end{eqnarray}
where $\phi_t = \phi \mod \pi/2$.
The base pixels have boundaries defined as
\begin{equation}
z > \frac{2}{3}, \quad \phi = k' \frac{\pi}{2},
\end{equation}
with $k'=0,1,2,3$

Similarly, for $2/3\leq z \leq -2/3$ the pixel boundaries can be found by setting $j=k
+s/2\pm (i-\nside)/2$ in Eq.~(\ref{eq:pixeqbeltphi}) and substituting into Eq.~(\ref{eq:pixeqbeltz})
\begin{equation}
z = \frac{2}{3} - \frac{4k}{3\nside} \pm \frac{8\phi}{3\pi}
\end{equation}

Using these pixel boundaries, one can easily  check by integration that each individual pixel has the
same surface area $\Opix$.

Table~\ref{table:resolution} summaries the number of pixels and resolution
available in \hpx. Since all pixels have the same surface area but slightly different
shape, the angular resolution is defined as 
\begin{eqnarray}
	\thetapix & \equiv & \sqrt{\Opix} \\
		  & = & \sqrt{\frac{3}{\pi}} \frac{3600}{1'} \frac{1}{\nside}
\end{eqnarray}


\section{Spherical Harmonic Transforms}

\begin{figure}[p]
\centerline{\psfig{figure=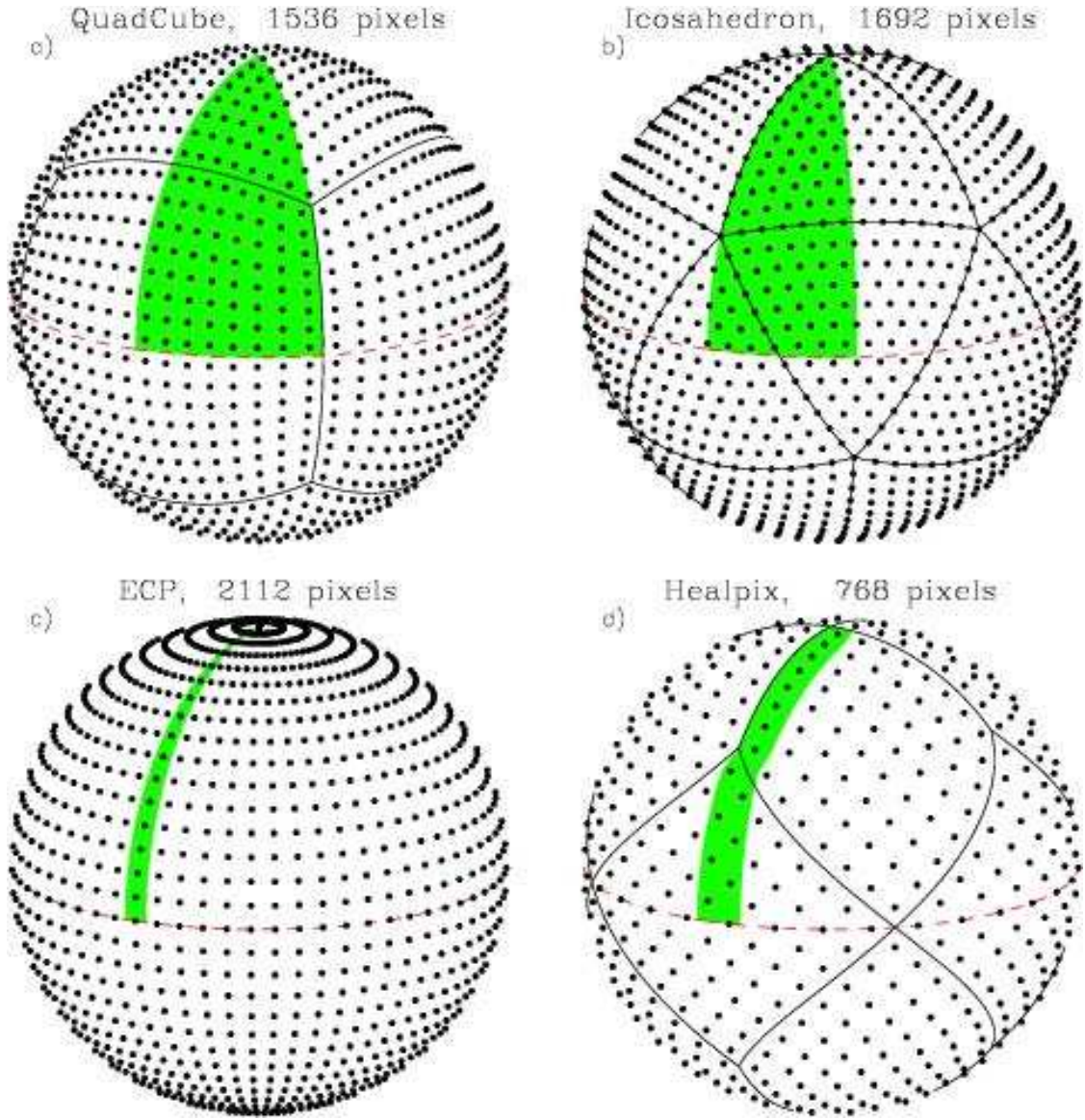,width=0.9\textwidth}}
  \caption{Comparison of HEALPix with other tessellations including
Quadrilateralized Spherical Cube (or QuadCube, used by NASA as data structure
for COBE mission products), icosahedral tessellation of the sphere, and
Equidistant Coordinate Partition, or the 'geographic grid.' The shaded areas illustrate
the subsets of all pixels on the sky for which the associated Legendre functions have to be computed 
in order to perform the spherical harmonic transforms. 
This plot demonstrates why the iso-latitude ECP and HEALPix points-sets support faster
computation of spherical harmonic transforms than the QuadCube, the icoshedral
grid, and any non iso-latitude construction.}
  \label{fig:tilings}
\end{figure}

The requirement of iso-latitudinal location of pixel centers was built into HEALPix in order for 
the grid to
support the fast discrete spherical harmonic transforms. The reason for the fast ($\tilde N_{pix}^{3/2}$)
scaling of the harmonic transform computational time with the size of the grid (or the grid resolution)
is entirely geometrical - the associated Legendre function components of spherical harmonics, which
can only be generated via slow recursions, have to be evaluated only once for each pixel ring.
For other grids, which are not iso-latitude constrainted, the extra computing time wasted on
non-optimal generation of the associated Legendre functions, typically results with
computational performance of order $\sim N_{pix}^2$. This geometrical aspect of 
discrete spherical transform computation is illustrated in Fig.\ref{fig:tilings},  
which compares HEALPix with other tessellations including
Quadrilateralized Spherical Cube (or QuadCube, used by NASA as data structure
for COBE mission products), icosahedral tessellation of the sphere, and
Equidistant Coordinate Partition, or the "geographic grid." This plot makes it visually
clear why the iso-latitude ECP and HEALPix points-sets support faster
computation of spherical harmonic transforms than the QuadCube, the icoshedral
grid, and any non iso-latitude construction.

\begin{figure}[ht]
   \includegraphics[width=0.9\textwidth,clip]{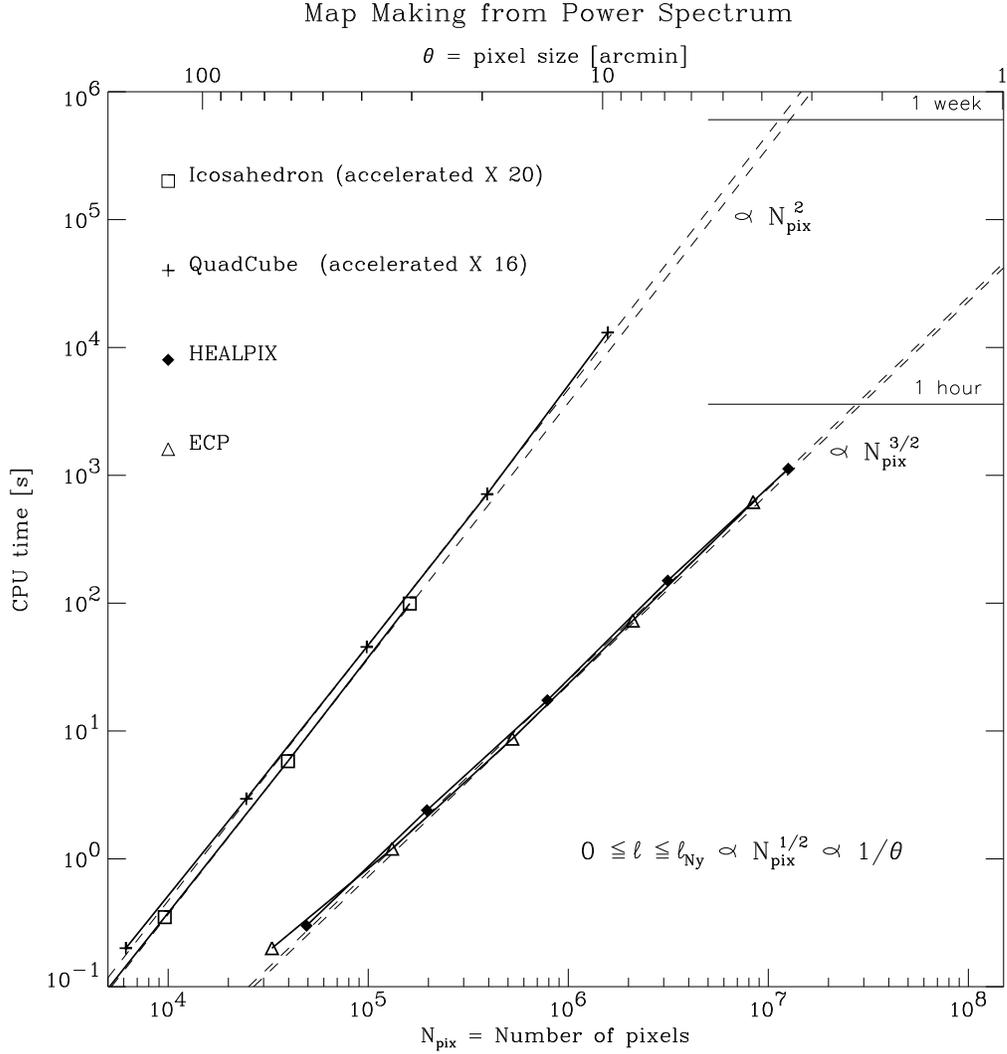}
  \caption{Illustration of the fundamental difference between computing speeds, which
can be achieved on iso-latitude and non iso-latitude point-sets. In order to be
able to perform the necessary computational work in support of the multi-million
pixel spherical data sets one has to resort to iso-latitude structures of
point-sets/sky maps, e.g. HEALPix Moreover, the future needs are already fairly
clear Ð measurements of the CMB polarization will require massively
multi-element arrays of detectors, and will produce data sets characterized by a
great multiplicity (of order of a few thousand) of sky maps. Since there are no
computationally faster methods than those already employed in HEALPix, and
global synthesis/analysis of a multimillion pixel map consumes about ~$10^3$s of a
standard serial machine CPU time, the necessary speed-up will have to be
achieved via optimized parallelization of the required computing.}
  \label{fig:timing}
\end {figure}

Figure (\ref{fig:timing}) demonstrates the fundamental difference between computing speeds, which
can be achieved on iso-latitude and non iso-latitude point-sets. In order to be
able to perform the necessary computational work in support of the multi-million
pixel spherical data sets one has to resort to iso-latitude structures of
point-sets/sky maps, e.g. HEALPix.  Moreover, the future needs are already fairly
clear -- measurements of the CMB polarization will require massively
multi-element arrays of detectors, and will produce data sets characterized by a
great multiplicity (of order of a few thousand) of sky maps. Since there are no
computationally faster methods than those already employed in HEALPix, and
global synthesis/analysis of a multimillion pixel map consumes about ~103s of a
standard serial machine CPU time, the necessary speed-up will have to be
achieved via optimized parallelization of the required computing.

A detailed description of implementation of the spherical harmonic transforms in the HEALPix software
package, and
the analysis of performance and accuracy thereof will given in a dedicated  publication.


\section{Summary}

The Hierarchical Equal Area iso-Latitude Pixelization, HEALPix, is a methodology of
discretization and fast numerical analysis and synthesis of functions or data
distributed on the sphere. HEALPix is an intermediate data-structural,
algorithmic, and functional layer between astronomical data, and the domain of
application of variety of science extraction tools. HEALPix as a sky map format
and associated set of analysis and visualization tools is already extensively
adopted as an interface between Information Technology and Space (and
suborbital) Science. This is manifested by applications of HEALPix by the
following projects: CMB experiments Boomerang (deBernardis et al, 2000, Ruhl et
al 2003), Archeops (Benoit et al, 2003ab), and TopHat, satellite mission WMAP
(WMAP2003), the forthcoming satellite mission Planck, the Sloan Digital Sky Survey,
and others. 


\begin{acknowledgements}
Our involvement with development, distribution, and support of HEALPix since 1997 would have been
impossible without support of a number of institutions and individuals. We are indebted to
Theoretical Astrophysics Center in Copenhagen,  Igor Novikov, and Per Rex Christensen;
to European Southern Observatory,  Peter Quinn, and Kevin Maguire; to MPA, Garching, and
Simon D.M. White;
to Caltech's Observational Cosmology Group, and Andrew Lange; to Caltech/IPAC, George Helou,
and Ken Ganga;
to JPL/Caltech, Center for Long Wavelength Astrophysics, and Charles R. Lawrence.
We are also grateful for all the positive feedback received from numerous HEALPix
users worldwide.
\end{acknowledgements}


\begin{table}[H]
\begin{tabular}{rrrrc}
\tableline
$k$ & $\nside=2^k$ & $ \Npix=12\nside^2 $ & $\thetapix=\Opix^{1/2}$ \\
\tableline
  0 & $     1 $ & $         12 $ & $        58.6^o $ \\
  1 & $     2 $ & $         48 $ & $        29.3^o $ \\
  2 & $     4 $ & $        192 $ & $        14.7^o $ \\
  3 & $     8 $ & $        768 $ & $        7.33^o $ \\
  4 & $    16 $ & $       3072 $ & $        3.66^o $ \\
  5 & $    32 $ & $      12288 $ & $        1.83^o $ \\
  6 & $    64 $ & $      49152 $ & $        55.0' $ \\
  7 & $   128 $ & $     196608 $ & $        27.5' $ \\
  8 & $   256 $ & $     786432 $ & $        13.7' $ \\
  9 & $   512 $ & $    3145728 $ & $        6.87' $ \\
 10 & $  1024 $ & $   12582912 $ & $        3.44' $ \\
 11 & $  2048 $ & $   50331648 $ & $        1.72' $ \\
 12 & $  4096 $ & $  201326592 $ & $        51.5'' $ \\
 13 & $  8192 $ & $  805306368 $ & $        25.8'' $ \\
\tableline
 14 & $  2^{14} $ & $   3.22\times 10^{ 9} $ & $       12.9'' $ \\
 15 & $  2^{15} $ & $   1.29\times 10^{10} $ & $       6.44'' $ \\
 16 & $  2^{16} $ & $   5.15\times 10^{10} $ & $       3.22'' $ \\
 17 & $  2^{17} $ & $   2.06\times 10^{11} $ & $       1.61'' $ \\
 $\vdots$ &  \vdots     &       \vdots              &     \vdots                 \\
 29 & $  2^{29} $ & $   3.46\times 10^{18} $ & $ {3.93\times 10^{-4}}'' $ \\
\tableline
\end{tabular}
\caption{Table of the number of pixels and pixel size accessible to
\hpx. The use of 32-bit signed integers for the pixel indexing currently
restrict the resolution accessible to $\nside \le 8192$. The use of 64-bit
signed integers will allow to reach $\nside = 2^{29}$.  }
\label{table:resolution}
\end{table}

\end{document}